\documentclass[article,twocolumn,showpacs,preprintnumbers,amsmath,amssymb]{revtex4}
\usepackage{graphicx}
\usepackage{dcolumn}
\usepackage{bm}
\usepackage{epsfig}

\begin{document}

\newcommand{\vk}{{\vec k}}
\newcommand{\vK}{{\vec K}} 
\newcommand{\vb}{{\vec b}} 
\newcommand{{\vp}}{{\vec p}} 
\newcommand{{\vq}}{{\vec q}} 
\newcommand{\vx}{{\vec x}}
\newcommand{\tr}{{{\rm Tr}}} 
\newcommand{\beq}{\begin{equation}}
\newcommand{\eeq}[1]{\label{#1} \end{equation}} 
\newcommand{\half}{{\textstyle \frac{1}{2} }}
\newcommand{\lton}{\mathrel{\lower.9ex 
\hbox{$\stackrel{\displaystyle <}{\sim}$}}} 
\newcommand{\gton}{\mathrel{\lower.9ex 
\hbox{$\stackrel{\displaystyle >}{\sim}$}}} 
\newcommand{\ee}{\end{equation}}
\newcommand{\bea}{\begin{eqnarray}} 
\newcommand{\eea}{\end{eqnarray}}
\newcommand{\beqar}{\begin{eqnarray}} 
\newcommand{\eeqar}[1]{\label{#1}\end{eqnarray}} 
\newcommand{\pleft}{\stackrel{\leftarrow}{\partial}}
\newcommand{\pright}{\stackrel{\rightarrow}{\partial}}

\begin{flushright}
\vskip .5cm
\end{flushright} \vspace{1cm}

\title{ Collisional dissociation of heavy mesons in dense QCD matter }

\author{Azfar Adil}%
\email{azfar@phys.columbia.edu}

\affiliation{Columbia University, 
Department of Physics and Astronomy,
New York, NY 10027, USA  }%

\affiliation{Frankfurt Institute for Advanced 
Studies (FIAS), 60438 Frankfurt am Main, Germany}

\author{Ivan Vitev}
\email{ivitev@lanl.gov}

\affiliation{  Los Alamos National Laboratory, Theoretical 
Division and Physics Division, Los Alamos, NM 87545, USA } %

\vspace*{1cm}

\begin{abstract}
In the framework of the reaction operator approach we calculate 
and resum the multiple elastic scattering of a fast $q \bar{q}$ 
system traversing dense nuclear matter. We derive the collisional
broadening of the meson's transverse momentum and the distortion 
of its intrinsic light cone wave function. The medium-induced 
dissociation probability of heavy mesons is shown to be sensitive 
to the opacity of the quark-gluon plasma and the time dependence of 
its formation and evolution. We solve the system of coupled rate 
equations that describe the competition between the fragmentation 
of $c$- and $b$-quarks and the QGP-induced dissociation of the 
$D$- and $B$-mesons to evaluate the quenching of heavy hadrons 
in nucleus-nucleus collisions. In contrast to previous results 
on heavy quark modification, this approach predicts suppression 
of $B$-mesons comparable to that of $D$-mesons at transverse 
momenta as low as $p_T \sim 10$~GeV. It allows for an 
improved description of the large attenuation of non-photonic 
electrons in central Au+Au reactions at RHIC.
\end{abstract}


\maketitle


\section{Introduction}

One of the important experimental signatures of the
Quark  Gluon Plasma (QGP) creation in heavy ion collisions 
is the  detailed suppression pattern of high transverse 
momentum  hadrons~\cite{Gyulassy:2003mc,Baier:2001yt}.  
Jet quenching for light mesons, such as $\pi$, $K$ and 
$\eta$, at the Relativistic Heavy Ion  Collider (RHIC) is 
well explained by radiative energy loss 
calculations~\cite{Adil:2004cn,Vitev:2005he}. In contrast,
recent data~\cite{Adler:2005xv,Ralf,Abelev:2006db,Alex} on the 
suppression of single non-photonic electrons  cannot  be 
reproduced   by  medium-induced  gluon  bremsstrahlung 
phenomenology in conjunction  with a physically 
reasonable set of QGP temperatures and 
densities~\cite{Djordjevic:2005db,Armesto:2005mz}. Collisional  
energy loss~\cite{Mustafa:2004dr,Peigne:2005rk,Adil:2006ei} 
has been proposed as a mechanism to reduce the discrepancy
between the experimental measurements and the perturbative 
QCD (PQCD) estimates of heavy flavor quenching. Still, up to 
transverse momenta $p_T \sim 10$~GeV partonic energy loss 
calculations under-predict the suppression of the non-photonic 
electrons by close to a factor of two~\cite{Wicks:2005gt}.  
The cause of this disagreement is readily identified with 
the small quenching of $B$-mesons, which dominate the 
high-$p_T$  single electron yields.  Similar results have 
been found in a Langevin 
approach to heavy quark diffusion where strong heavy quark 
interactions with $D$- and $B$-resonances near the 
phase transition have been 
assumed~\cite{vanHees:2004gq,Moore:2004tg}.

The model for predicting the quenched hadron spectra 
in $A+A$ collisions, where final-state in-medium effects 
are  important, has been very simple: it is assumed
that the hard jet hadronizes in vacuum,
having fully traversed the region of dense nuclear matter, 
$L_T^{\rm QGP} \leq 6$~fm, 
and lost energy via radiative and collisional processes.
Let us examine the validity of this assumption for 
different species of final-state 
partons and decay hadrons. If a parton of non-zero mass 
$m_Q$ fragments into a massive hadron $m_h$ and a secondary  
light parton, 
\begin{eqnarray} 
\left[p^+, \frac{m_Q^2}{2p^+}, {\bf 0} \right]
& \rightarrow & \left[zp^+, \frac{ {\bf k}^2 +  m_h^2}{2zp^+}, 
{\bf k} \right]  \nonumber \\
&& + 
 \left[(1-z)p^+, \frac{ {\bf k}^2 }{2(1-z)p^+}, 
-{\bf k} \right] \;.  
\label{fragtime}
\end{eqnarray} 
In Eq.~(\ref{fragtime}), $p^+$ is the large light cone momentum 
of the parton, $z = k^+/p^+$ and $|{\bf k}| \sim \Lambda_{QCD}
\sim 200$~MeV  is the deviation from collinearity.
We can estimate from the uncertainty principle for the variable
conjugate to the non-conserved light cone momentum component 
$\Delta p^- = (p^-)_f - (p^-)_i $:   
\begin{eqnarray} 
\Delta y^+ &\simeq& \frac{1}{\Delta p^-} \; = \; 
\frac{2z(1-z)p^+}{ {\bf k}^2 + (1-z)m_h^2 - z(1-z)m_Q^2   } \;. \quad
\label{tfrag}
\end{eqnarray} 
With $\Delta y^+ =  \tau_{\rm form} +  z_{\rm form}$,  
$z_{\rm form} = \beta_Q   \tau_{\rm form}$ being the formation 
distance,  the formation  time reads:
\begin{eqnarray} 
&& \tau_{\rm form} = \frac{ \Delta y^+}{ 1+\beta_Q} \;, \quad 
\beta_Q = \frac{p_Q}{E_Q} \;. 
\label{tform}
\end{eqnarray} 
Clearly, the higher the energy, or $p_T$, and the lighter the hadron - 
the better the perturbative treatment of partonic energy loss prior 
to hadronization will be~\cite{Wang:2003aw}. 
For a $p_T = 10$~GeV pion at mid-rapidity
$\tau_{\rm form} \approx 25$~fm $\gg L_T^{\rm QGP}$, consistent with 
the jet quenching assumptions and the robust description of the nuclear 
modification factor $R_{AA}(p_T)$ versus $\sqrt{s_{NN}}$ and 
centrality~\cite{Gyulassy:2003mc,Adil:2004cn,Vitev:2005he}.  
In contrast, $B$- and $D$-mesons of the same $p_T$ have 
formation times $\tau_{\rm form} \approx 0.4$~fm and $1.6$~fm, respectively, 
$\ll L_T^{\rm QGP}$. For the purpose of these estimates we 
used $\bar{z}(\pi) =0.7$,   $\bar{z}(D) =0.82$ and $\bar{z}(B) =0.92$
and the exact hadron masses in Eq.~(\ref{tfrag}). 
Therefore, at the finite $p_T$ range accessible 
at RHIC and  LHC a conceptually different approach to the 
description of $D$- and $B$-meson quenching in A+A collisions 
is required, when compared to light hadrons. In the future in 
the limit of high transverse momenta, $p_T(D) > 10$~GeV and  
$p_T(B) > 30$~GeV, improved PQCD description of heavy flavor 
suppression should seek to also incorporate the traditional partonic 
energy loss mechanisms.     

This Letter summarizes key results from our calculation of the 
collisional dissociation of heavy mesons in the quark-gluon plasma. 
In Section II we determine the baseline cross sections for heavy 
flavor production on the example of the identified $D^0$, $D^+$ and $B^+$
spectra from the Tevatron. The light cone wave functions for $D$- and 
$B$-mesons are constructed in Section III.  In Section IV we use the 
Gyulassy-Levai-Vitev (GLV) reaction operator approach to derive 
the collisional broadening of the propagating $q\bar{q}$ system 
and evaluate the survival and  dissociation probabilities of heavy 
mesons  as a function of the cumulative transverse momentum transfer
from the medium. In Section V we  solve the system of coupled rate 
equations for the fragmentation into and  dissociation of heavy mesons. 
Phenomenological comparison to RHIC data on the quenching of 
non-photonic electrons is presented. Our conclusions are given in 
Section VI.

\section{$D$- and $B$-meson production in p+p collisions}

Single inclusive hadron production can be calculated to lowest 
order in the PQCD collinear factorization approach as follows:
\begin{eqnarray}
\frac{ d\sigma^{H}_{NN} }{ dy  d^2p_{T}  } 
& = &  K \sum_{abcd} \int_{\cal D}  dy_d   d z  \; 
 \frac{1}{z^2} D_{H/c}(z) 
\, \alpha_s^2(\mu_r) |\overline {M}_{ab\rightarrow cd}|^2 
   \nonumber \\
&& \times  \frac{\phi_{a/N}({x}_a,\mu_f) 
\phi_{b/N}({x}_b,\mu_f) }{{x}_a{x}_b{S}^2}   \;.
\label{single} 
\end{eqnarray}
In  Eq.~(\ref{single})  $S = (p_A + p_B)^2$  is the 
squared center of mass energy of the hadronic collision,  
$x_a = p_a^+/{p_A}^+$,   $x_b = p_b^-/{p_B}^-$  are 
the light cone momentum fractions of the incoming partons and 
$ \phi_{a/N}(x_{a}), \phi_{b/N}(x_{b})$ are the parton distribution 
functions~\cite{Pumplin:2002vw}.  In our calculation the 
factorization and renormalization  scales are set to  
$\mu_f^2 = \mu_r^2 =  {p_{Tc}^2+m_Q^2}$. The running of the strong
coupling constant is taken to lowest order 
with $N_f = 4$ active quark flavors and we use $m_c = 1.3$~GeV, 
$m_b = 4.5$~GeV. Fragmentation functions 
$D_{H/c}(z)$ of heavy quarks into mesons, derived 
in~\cite{Braaten:1994bz,Cheung:1995ye}, are used.  At lowest
order we include $Q+g \rightarrow Q+g$, $Q+q \rightarrow Q+q$ and 
$Q+\bar{q} \rightarrow Q+\bar{q}$ processes, using standard heavy quark
PDFs, in addition to gluon fusion and light quark annihilation. 
We find that these processes give a dominant contribution to the 
single inclusive $D$- and $B$- meson cross section. They ensure faster 
convergence of the perturbation series and small next-to-leading 
order $K$-factors. Further details of the perturbative calculation are 
given in~\cite{Vitev:2006bi}.

\begin{figure}[t]
\begin{center} 
  \psfig{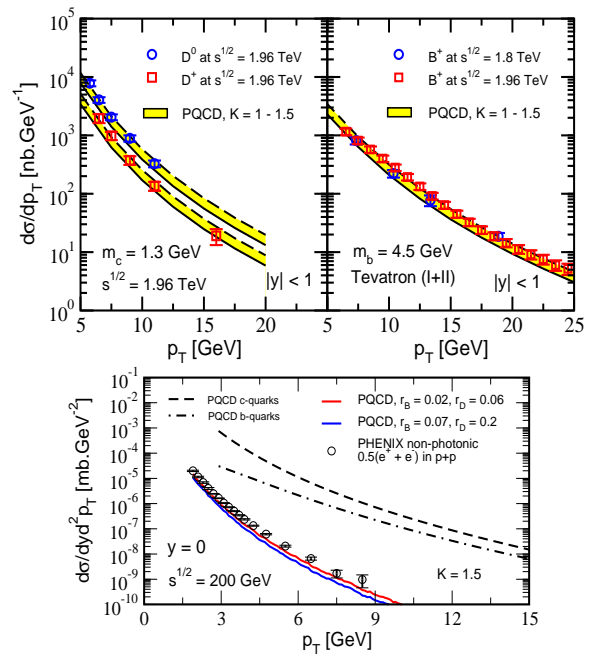}
\caption{Perturbative QCD calculation~\cite{Vitev:2006bi} of the 
$D$- and $B$-meson production cross sections at $\sqrt{s} = 1.96$~TeV,
$|y| \leq 1$. 
The top panels show comparison to $D^0$ and $D^+$ experimental measurements
at the Tevatron~\cite{Acosta:2003ax} and $B^+$ 
data~\cite{Acosta:2001rz,Acosta:2004yw}. In the bottom panel, the
PQCD calculation confronts the  $\sqrt{s} = 200$~GeV non-photonic 
electron data by PHEHIX at RHIC~\cite{Adare:2006hc}. The corresponding
c- and b-quark cross sections are also shown.}
\label{tevbase}
\end{center} 
\end{figure}

Figure~\ref{tevbase} shows comparison of the PQCD single inclusive 
spectra, Eq.~(\ref{single}), to Tevatron measurements at 
$\sqrt{s} = 1.8$~GeV and 
1.96~GeV~\cite{Acosta:2003ax,Acosta:2001rz,Acosta:2004yw}.  
We observe that  a similar description of the $B^+$ cross 
section can be achieved when compared to $D^0$ and $D^+$ cross 
sections and note that $b$-quark fragmentation functions are
harder $(r = 0.07)$  than the $c$-quark fragmentation 
functions $(r = 0.20)$. Here, $r$ is the parameter that 
determines the hardness of these heavy quark fragmentation 
functions~\cite{Braaten:1994bz,Cheung:1995ye} and 
$r_B/r_D \sim m_c/m_b$.
We also anticipate that at $p_T \leq m_b$ large corrections 
will arise to the LO perturbative calculation of $B$-mesons. 
However, in this region $d\sigma^B/dydp_T < d \sigma^D/dydp_T$  
has a smaller contribution to the heavy meson production rate.  
We use the same choice of scales and $K=1.5$ as the baseline 
for studying nuclear effects on heavy flavor production at the smaller 
$\sqrt{s_{NN}} = 200$~GeV at RHIC and the larger 
$\sqrt{s_{NN}} = 5.5$~TeV at  LHC. We note that  
$K$-factors cancel in the nuclear  modification ratio 
$R_{AA}(p_T)$. For baryons, which constitute 
$\sim 10\%$ of the total $c$- and $b$- quark decays, we use softer 
fragmentation functions with $r_{\rm baryon} = 1.5 \, r_{\rm meson}$.
In the bottom panel of Fig.~\ref{tevbase}  results for  
the electrons, $0.5(e^+ + e^-)$, from the semi-leptonic decays
of heavy flavor, evaluated in LO PQCD  in p+p collisions 
at RHIC, are shown.  Data is from PHENIX~\cite{Adare:2006hc}.
Harder  fragmentation functions, $r_B = 0.02,  r_D = 0.06$, have 
also been used for direct comparison to alternative  calculations.


\section{Light cone wave functions}

In order to calculate the effect of the QGP medium on a meson traveling 
through, we need a model for the meson state in terms of 
its partonic degrees of freedom $(x_i, {\bf k}_i)$.  Here 
$x_i = k^+_i/P^+, \; \sum_{i =1}^{n\geq 2} x_i = 1 $ are the light cone
momentum fractions and ${\bf k}_i, \;  \sum_{i=1}^{n\geq 2} 
{\bf k}_i = {\bf K}$, are the internal parton transverse 
momenta.  To determine the typical momenta  $\langle {\bf k}_i^2 
\rangle$,   we examine the numerical results of potential model 
calculations of $D$- and $B$-meson mass spectra and decay 
widths~\cite{Avila:1994vi}. Solving the Dirac equation for the 
relativistic light quark in the potential of the heavy 
quark $V = -\xi /r + br$, one can achieve a very good description 
of the lowest-lying and excited heavy meson states. The radial 
wave function of the light quark $\rho(r) = r^2 |\psi(r)|^2$ is 
found to have its maximum at $a_0 = 2-3$~GeV$^{-1}$~\cite{Avila:1999aj}.    
Fourier transforming a Gaussian distribution, which features the 
same maximum of the radial wave function $\rho(r)$,  we determine 
the momentum distribution: 
$\psi(r) \sim e^{r^2/(2a_0^2)}  \; \rightarrow  \; 
\psi(k) \sim e^{k^2a_0^2/2}$.  
Evaluation of the mean transverse momentum squared is straightforward:
\beqar  
\langle {\bf k}^2 \rangle &=& N_k^2 \; 2 \pi \int_{-1}^{1} 
d \cos(\theta)  \, cos^2(\theta) \nonumber \\ 
&& \times \, \int_0^\infty k^2\, k\,  dk  
 \; e^{k^2 a_0^2}  = \frac{1}{2a_0^2} \;,     
\eeqar{kt2}  
where $N_k^2$ ensures the proper normalization of the 
distribution in momentum space. Taking 
$a_0 = 2.5$~GeV$^{-1}$ for the light parton we obtain 
$\langle {\bf k}^2 \rangle$  =  $0.08$~GeV$^2$.

To determine the longitudinal momentum fractions of the 
quarks in a boosted heavy meson we express the  state 
in terms of its multi-parton Fock components as 
follows~\cite{Brodsky:2000ii,Brodsky:2001wx}:
\begin{eqnarray}
&& |\Psi_{M},P^{+},{\bf P} \rangle = \sum_{n \geq 2} 
\int \prod_{i=1}^{n} 
\frac{dx_{i}d^{2}{\bf k}_{i}}{\sqrt{(2\pi)^3} \sqrt{2x_{i}}}
\delta\left(\sum_{i=1}^{n} x_{i}  - 1\right)
\nonumber \\
&& \times \,  \delta^{2} \left( \sum_{i=1}^{n} 
{\bf k}_{i} \right)
 \, \psi({\bf k}_{i},x_{i})| n;
{\bf k}_{i}+x_{i}{\bf P}, x_{i}P^{+} \rangle \;,
\label{brodswave}
\end{eqnarray}
where the light cone wave functions $\psi({\bf k}_{i},x_{i})$ of the 
partons in the meson are universal. Note that these do not depend 
on the external meson momenta, $P^+, {\bf P}$. We have chosen the 
following normalization for the single parton states: 
\begin{eqnarray}
\langle x^\prime P^+, {\bf k}^\prime ; 1  | 1;
{\bf k}, x P^+ \rangle  = (2\pi)^3 2 x \,\delta(x - x^\prime)
\delta^2({\bf k}- {\bf k}^\prime ) \;. \quad 
\label{norm}
\end{eqnarray}

In our calculations we take the lowest 
lying $n=2$ Fock component. A heavy meson moving in the 
positive light cone direction  will then be described by 
\begin{eqnarray}
|\psi({\bf K},\Delta {\bf k};x,m_{1},m_{2})|^{2} &=& {\rm Norm}^2 
e^{-\frac{ \Delta {\bf k}^{2}+4m_{1}^{2}(1-x)+4m_{2}^{2}x}
{4x(1-x)\Lambda^{2}}} \nonumber \\
&&\times  \delta^{2}({\bf K}) \; , \;
\label{wavesq}
\end{eqnarray}
where the overall large light cone momentum has been integrated out. 
In Eq. (\ref{wavesq}) ${\bf K} = {\bf k}_{1}+{\bf k}_{2}$ 
and $\Delta {\bf k}={\bf k}_{1}-{\bf k}_{2}$.  We assume here that 
$x$ is the momentum fraction carried by the  heavy quark $Q$ and
use $m_{u}=m_{d}=0.005$ GeV, $m_{c}=1.3$ GeV and $m_{b}=4.5$ GeV.
The light cone wave function satisfies the requirement that 
the maximum in the longitudinal momentum density distribution 
is achieved when
the constituent partons of the meson  are at the same rapidity
or, equivalently, 
$m_{T_1}/x_1 = m_{T_2}/x_2$~\cite{Brodsky:2000ii,Brodsky:2001wx}. 
Fixing the light quark momentum fraction squared   
$ \langle (\Delta{\bf k}/2)^2 \rangle $  to the value 
obtained in Eq.~(\ref{kt2}), we can determine  $\Lambda = 0.735$~GeV 
and $\Lambda = 1.055$~GeV for $D$-mesons and $B$-mesons, respectively.

\begin{figure}[t]
\begin{center} 
\psfig{file=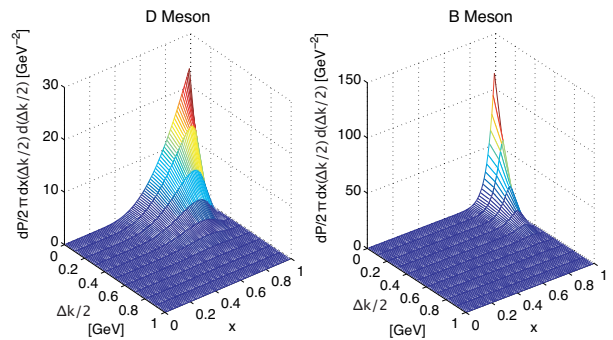,height=1.8in,width=3.6in,angle=0}
\caption{Surface plots of the momentum density $|\psi(x,\Delta k /2)|^2$  
for the lowest lying $Q \bar{q}$ (or $\bar{Q} q$)  Fock component
of the $D$-  and $B$-meson states.}
\label{sqwavsurf}
\end{center} 
\end{figure}

Figure \ref{sqwavsurf} show the momentum density distribution of the 
heavy quark inside the heavy mesons from Eq.~(\ref{wavesq}) plotted 
versus the light cone momentum fraction $x$ and its half relative 
transverse momentum $ \Delta k  /2 = | \Delta {\bf k} / 2 |$.
The following normalization, 
 $\int d^{2} \Delta {\bf k} dx\; |\psi(x,\Delta {\bf k})|^{2} = 1$,
was used. We note the momentum distribution is fairly narrow in 
$  \Delta  k  / 2 $ and momentum transfers from the medium with 
$\mu \geq 1$~GeV~\cite{Gyulassy:2003mc,Adil:2004cn,Vitev:2005he} at the 
initial stages of the evolution of the QGP density may easily dissociate 
the heavy mesons if they tend to form early $\ll L_T^{QGP}$, see 
Eqs.~(\ref{fragtime}) and (\ref{tform}). Integrating over $\Delta {\bf k}$ 
we obtain the parton distribution function of the heavy quark inside 
the meson:
\begin{eqnarray}
\phi_{Q/M}(x)& =& \int d^{2}\Delta {\bf k}d^{2}{\bf K}\, 
|\psi({\bf K},\Delta {\bf k};x,m_{1},m_{2})|^{2} \nonumber \\
&=& {\rm Norm}^2 4\pi x(1-x)\Lambda^{2}
e^{-\frac{m_{1}^{2}(1-x)+m_{2}^{2}x}{x(1-x)\Lambda^{2}}} \;.
\label{phiofx}
\end{eqnarray}
We note that the heavy $c$- and $b$-quark distributions, Eq.~(\ref{phiofx}), 
closely resemble in shape the fragmentation functions 
$D_{H/Q}(z)$~\cite{Braaten:1994bz,Cheung:1995ye} and peak toward
larger values of $x$ with increasing heavy quark mass.


\section{Collisional dissociation of heavy mesons}

The GLV reaction operator formalism~\cite{Gyulassy:2000fs,Gyulassy:2000er}  
was developed for calculating the induced radiative energy 
loss of hard quarks or gluons when they pass through a dense 
medium. In this approach, the multi-parton dynamics is 
described by a series expansion in 
$\chi = \int_0^{L_T^{QGP}} \sigma_{el}(z) \rho(z) dz =  L_T^{QGP}/ 
\langle \lambda  \rangle $, 
the mean number of interactions that a fast projectile undergoes 
along its trajectory.  Each interaction is represented by a 
reaction operator that summarizes the unitarized basic scattering 
between the propagating system and the medium. The summation to all 
orders in opacity, $\chi $,  is achieved by a recursion of the 
reaction operator. For the  case of collisional interactions of individual
fast partons in dense QCD matter, their diffusion in transverse 
momentum space has been derived to leading power~\cite{Gyulassy:2002yv} 
and leading power corrections~\cite{Qiu:2003pm}.

\begin{figure}[t]
\begin{center} 
\psfig{file=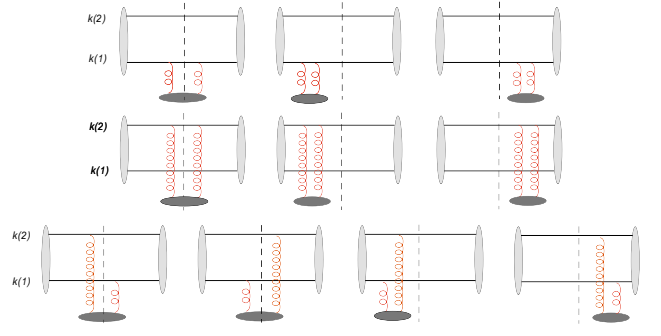,height=1.7in,width=3.3in,angle=0}
\caption{Forward cut diagrams representing one collisional interaction 
of a $q\bar{q}$  system with momenta $k_{1}$ and $k_{2}$ 
propagating through dense nuclear medium.}
\label{diags}
\end{center} 
\end{figure}

The transverse momentum transfer ${\bf q}_n$ at position $n$ 
to a fast parton when it scatters on the soft constituents of 
the medium is distributed  according to the normalized 
differential cross section  
\begin{equation}
\frac{1}{\sigma_{el}(n)}\frac{d \sigma_{el}(n)}{d^2 {\bf q}_n}  
= \frac{\mu^2(n)}{\pi({\bf q}_n^2+\mu^2(n))^2} \; .
\label{GWmodel}
\end{equation}
In Eq.~(\ref{GWmodel}), $\mu(n)=gT(n)$ is the thermally 
generated Debye screening mass,  $\sigma_{el}(n) 
\approx  C_{ij}   2 \pi \alpha_s^2/\mu^2(n)$ with
$C_{ij} = 9/4, 1, 4/9$ for $gg, qg, qq$, respectively,  
and the mean free path  $\lambda (n)=1/(\sigma_{el}(n)\rho(n))$.  
Two momentum transfers are necessary to build one power of the 
elastic scattering cross section in Eq.~(\ref{GWmodel}),  
allowing for three $t = \infty$ on-shell cuts in the forward
scattering Feynman diagrams, see Fig.~\ref{diags}. Momentum 
flow in the legs of the direct- and virtual-interaction terms 
is constrained 
as follows:
\begin{equation}
{\rm Dir.} \sim \delta^2({\bf q}_n - {\bf q}_n^\prime) \;, 
\quad {\rm Vir.} \sim - \frac{1}{2} 
\delta^2({\bf q}_n + {\bf q}_n^\prime) \;. 
\label{constr}
\end{equation}
For further details on the derivation of this formalism, 
see~\cite{Gyulassy:2000fs,Gyulassy:2000er,Gyulassy:2002yv,Qiu:2003pm}.

The diagrams that are relevant to a single in-medium interaction of the 
quark-antiquark system are shown in  Fig.~\ref{diags}.  We work in terms 
of the momenta ${\bf K}$ and $ \Delta {\bf k}$, defined in 
the previous Section. If the momentum distribution of the $q\bar{q}$
system that has undergone $n$ scatterings is 
$\propto | M_{n}^\star ({\bf K}, \Delta {\bf k})  
M_{n} ({\bf K}, \Delta {\bf k}) | $,
it can be related to the momentum density prior to the last collision 
as follows:
\begin{eqnarray}  
&& \hspace*{-0.4cm} M_{n-1}^\star({\bf K},  \Delta {\bf k})  \left[  
e^{-{\bf q}_{n} \cdot \stackrel{\leftarrow}{{\nabla}_{{\bf K}}}}
e^{-{\bf q}_{n} \cdot \stackrel{\rightarrow}{{\nabla}_{{\bf K}}}}
\otimes \left(  
e^{-{\bf q}_{n} \cdot \stackrel{\leftarrow}{{\nabla}_{ \Delta {\bf k}}}}  
e^{-{\bf q}_{n} \cdot \stackrel{\rightarrow}{{\nabla}_{ \Delta {\bf k}}}}
\right. \right.   \nonumber \\ 
&& \hspace*{-0.4cm}  + \;  
e^{+{\bf q}_{n} \cdot \stackrel{\leftarrow}{{\nabla}_{\Delta {\bf k}}}}  
e^{+{\bf q}_{n} \cdot \stackrel{\rightarrow}{{\nabla}_{\Delta {\bf k}}}}
+ 
e^{-{\bf q}_{n} \cdot \stackrel{\leftarrow}{{\nabla}_{\Delta {\bf k}}}}  
e^{+{\bf q}_{n} \cdot \stackrel{\rightarrow}{{\nabla}_{\Delta {\bf k}}}}
 \nonumber \\ 
&&\hspace*{-0.4cm}  +  \left. 
e^{+{\bf q}_{n} \cdot \stackrel{\leftarrow}{{\nabla}_{\Delta {\bf k}}}}  
e^{-{\bf q}_{n} \cdot \stackrel{\rightarrow}{{\nabla}_{\Delta {\bf k}}}}
\right)  - \hat{\bf 1} \otimes \left( 
2 \; \hat{1} 
+ 
e^{-2{\bf q}_{n} \cdot \stackrel{\leftarrow}{{\nabla}_{\Delta {\bf k}}}}  
e^{+{\bf 0}_{n} \cdot \stackrel{\rightarrow}{{\nabla}_{\Delta {\bf k}}}}
  \right. \nonumber \\ 
&& 
\hspace*{-0.4cm}  
\left. \left.  +\;
e^{+{\bf 0}_{n} \cdot \stackrel{\leftarrow}{{\nabla}_{\Delta {\bf k}}}}  
e^{-2{\bf q}_{n} \cdot \stackrel{\rightarrow}{{\nabla}_{\Delta {\bf k}}}}
 \right)  \right]  M_{n-1} ({\bf K}, \Delta {\bf k}) \;.
\label{shiftn}
\end{eqnarray}  
Here $e^{-{\bf q}_{n} \cdot 
{{\nabla}_{ {\bf K}, \Delta {\bf k}} } } $ are the momentum shift 
operators~\cite{Gyulassy:2000er}.   
In Eq.~(\ref{shiftn}) we 
assume that the overlap between the amplitude and its conjugate 
varies slowly with  ${\bf q}_{n}$ and  symmetrize the momentum shifts  
in the last two virtual terms 
around  ${\bf q}_{n} = 0$. This allows us to write at the level of 
the momentum distributions (squared amplitudes):
\begin{eqnarray}  
&&  |M_n ({\bf K},  \Delta {\bf k})|^{2}  \propto  
\left[ 2 \left( 
e^{-{\bf q}_{n} \cdot \stackrel{\rightarrow}{{\nabla}_{{\bf K}}}}
- \hat{\bf 1} \right) \cosh  \left(  e^{-{\bf q}_{n} 
\cdot \stackrel{\rightarrow}{{\nabla}_{ {\Delta \bf k}}}}    \right)
\right. \nonumber \\
&& \left . + 2  \left(   e^{-{\bf q}_{n} \cdot 
\stackrel{\rightarrow}{{\nabla}_{\Delta {\bf k}}}}                 
  - \hat{1}   \right) 
 \right]  |M_{n-1} ({\bf K}, \Delta {\bf k})|^{2} \;.
\label{shiftn1}
\end{eqnarray}  
In  Eq.~(\ref{shiftn1}) we also used the symmetry of the 
momentum transfer distribution, Eq.~(\ref{GWmodel}), relative
to the transformation ${\bf q}_{n}  \rightarrow - {\bf q}_{n} $.
The basic step in Eq.~(\ref{shiftn1}) allows us to resum 
the interactions to all orders in opacity and relate 
the final momentum distribution of the evolved $q\bar{q}$ system 
to the one in Eq.~(\ref{wavesq}):
\begin{eqnarray}
|\psi _{f}({\bf K}, \Delta {\bf k})|^{2}
& = &  \sum_{n=0}^{\infty} \frac{2 ^{n}\chi^n}{n!} 
\int  \prod_{i=1}^n   d^2 {\bf q}_{i }   \frac{1}{\sigma_{el}} 
\frac{d\sigma_{el} }{d^2{\bf q}_{i}} \,
\nonumber \\ &&  \hspace*{-1cm}
\times \left[  \left( e^{-{\bf q}_{n} \cdot 
\stackrel{\rightarrow}{{\nabla}_{{\bf K}}}}
- \hat{\bf 1} \right)  
\cosh  \left(  -{\bf q}_{n} 
\cdot \stackrel{\rightarrow}{{\nabla}_{ \Delta {\bf k}}}    \right)
 \right. \nonumber \\ &&  
\hspace*{-1cm}  \left. +  \left(   e^{-{\bf q}_{n} \cdot 
\stackrel{\rightarrow}{{\nabla}_{ \Delta {\bf k}}}}                 
  - \hat{1}  \right)  
 \right]  \,     |\psi_{0}({\bf K},\Delta {\bf k})|^{2}  \; . 
\label{fulliter}
\end{eqnarray}

An approximate closed form for Eq.~(\ref{fulliter}) can be 
obtained by Fourier transforming to the impact parameter space
$({\bf B},{\bf b})$ conjugate to  $({\bf K},\Delta{\bf k})$:
\beqar
|\widetilde{\psi_{f}} ({\bf B},{\bf b}) |^{2}
&=& |\widetilde{\psi_{0}}({\bf B},{\bf b})|^{2}
\; \exp \left[ 2 \chi \int d^2 {\bf q} \, \frac{1}{\sigma_{el}} 
\frac{d\sigma_{el}}{d^2{\bf q} }   \right. \nonumber \\ 
&&
\hspace*{-2cm} 
\times \left(\, \left( e^{i{\bf q} \cdot {\bf B} } -1\right)\cos 
\left( {\bf q} \cdot {\bf b} \right)+
\left( e^{i{\bf q} \cdot {\bf b} }-1\right) \, \right) \bigg] \;.  
\eeqar{bspace}
A closed form for the integral in Eq.~(\ref{bspace}) does not exist 
even for simple forms of the differential  elastic scattering cross 
section.  The cosine term couples the broadening of the total
momentum ${\bf K}$ of the quark-antiquark system to the distortion 
of the light cone wave function in  $\Delta {\bf k}$.
We can still calculate the final momentum distribution
by considering  only the leading effect of  the coupling: 
$\cos({\bf q} \cdot {\bf b})=1- ({\bf q} \cdot 
{\bf b})^{2}/2 +\cdots $. 
Next, we perform the integrals over the Yukawa potential.
With $q = |{\bf q}|$,
\begin{eqnarray}
\label{ctb1}
&& \int d^2 {\bf q}   \frac{ \mu^2}{\pi (\mu^2+q^2)^2} e^{iqb\cos(\phi)}
  =  b \mu \, K_1(b \mu) \;.
\end{eqnarray}
The $b \rightarrow  B$ contribution without interference yields a 
result similar to Eq.~(\ref{ctb1}) and the second term in the expansion
of the cosine gives 
\beqar
&& - \frac{\mu^{2}b^{2}}{2\pi} 
\int_0^{\frac{1}{B}} d^{2}{\bf q}  \, 
 \frac{q^2}{(\mu^2+q^2)^2} 
  \left(e^{i q B \cos(\phi)} -1 \right) \cos^{2}(\phi) 
\nonumber \\[1ex]
&&=  \frac{3b^{2}\mu^{2}}{32}+\frac{3 b^{2}\mu^2 B^{2}
\mu^{2}}{32}(1+2\log(B^{2}\mu^{2})) \; .
\eeqar{contribs}   
In Eq.~(\ref{contribs}) we have taken $B$ and $b$ in the same
direction and chosen $1/B$ as the upper limit of the $q$ integral
to obtain an estimate for the upper limit of the interference
contribution.
The key to evaluating the average ${\bf K}$- and 
$\Delta {\bf k}$-broadening  of the partons is the small 
$b \mu$ expansion in  Eqs.~(\ref{ctb1}) and (\ref{contribs}):
\begin{eqnarray} 
b \mu \, K_1(b \mu) &=& 
1- \frac{b^2 \mu^2}{2}\left[  \ln \left(\frac{2e^{-\gamma_E}}{b \mu} 
\right) + \frac{1}{2} \right]  \nonumber \\  && 
+ {\cal O}(b^4 \mu^4) \;.
\end{eqnarray} 
Note that the leading correction that arises from 
Eq.~(\ref{contribs}), $3b^2\mu^2/16$, is small. Keeping terms 
$ \propto \chi \mu^2 \xi$, where $\xi = 
\ln ( 2e^{-\gamma_E}/(b \mu )) + 1/2  - 3/16 \geq {\cal O}(1)$, and 
incorporating 
only the leading effect of the coupling term in Eq.~(\ref{contribs}) 
we find:
\beq
|\widetilde{\psi_{f}} ({\bf B},{\bf b}) |^{2}
=  |\widetilde{\psi_{0}}({\bf B},{\bf b}) |^{2} 
e^{- b^2 (\chi \mu^{2}\xi )}e^{- B^2 (\chi \mu^{2}\xi ) } .
\eeq{bspace-approx} 
We treat $\xi$ as approximately constant when compared to
the power behavior of $b^2$ and $B^{2}$ when Fourier 
transforming Eq.~(\ref{bspace-approx})
back to momentum space:
\begin{eqnarray}
&& \hspace*{-0.3cm} |\psi_{f}( {\bf K},\Delta {\bf k} )|^{2} 
= \int \frac{d^{2}{\bf b}}
{(2\pi)^{2}}\frac{d^{2}{\bf B}}{(2\pi)^{2}} \,
e^{-i {\bf b}\cdot \Delta {\bf k}}\,  e^{-i {\bf B}\cdot{\bf K}} \, 
e^{- B^2 (\chi \mu^{2} \xi) } \nonumber \\
&& {\hspace*{0.5cm}}\,\times \,
e^{-b^{2} (\chi \mu^{2}\xi )} \; \left( {\rm Norm^2} 
\,  4\pi\,  x(1-x) \Lambda^2  e^{-b^2 ( x(1-x)\Lambda^{2} ) } \right.
\nonumber \\
&& \left. \hspace*{3cm}
\times \,e^{-\frac{ m_{1}^{2}(1-x)+m_{2}^{2}x}
{x(1-x)\Lambda^{2}}}  \right) 
\nonumber \\
&& \hspace*{0.5cm}
= \left[ \frac{e^{-\frac{{\bf K}^{2}}{4\chi \mu^{2}\xi}}}
{4\pi \chi \mu^{2}\xi} \right]
\left[  {\rm Norm}^2 \,
 \frac{ x(1-x) \Lambda^2 }
{ \chi \mu^{2}\xi+x(1-x)\Lambda^{2} } 
\right. \nonumber \\
 && \hspace*{1.cm} \left. \times \;e^{-\frac{\Delta {\bf k}^{2}}
{4(\chi \mu^{2}\xi+x(1-x)\Lambda^{2})}}
\, e^{-\frac{ m_{1}^{2}(1-x)+m_{2}^{2}x}
{x(1-x)\Lambda^{2}}}  \right] \;.
\label{wavesqfin}
\end{eqnarray}   
In providing a physics interpretation to Eq.~(\ref{wavesqfin}),
we note that the first part of our result represent the broadening
in the momentum distribution of the meson itself, 
$\langle {\bf K}^2 \rangle = 4\chi \mu^{2} \xi $. It is twice the size
of the broadening for an individual parton~\cite{Qiu:2003pm}.
The second part, critically important for this work, represents 
the distortion in the intrinsic momentum distribution of the 
quarks, which leads to the meson decay.

We first integrate out the distribution in ${\bf K}$. The small 
acoplanarity that may arise in heavy meson or non-photonic electron
triggered correlations~\cite{Vitev:2006bi,Vitev:2003xu} is  
neglected in this work. Since our wave functions are real, they 
are given by the square root of the $\Delta {\bf k}$ and $x$ momentum 
distributions, Eqs.~(\ref{wavesq})  and (\ref{wavesqfin}).
In our calculation the initial-state $\psi_0(\Delta {\bf k}, x) = 
 \psi_M(\Delta {\bf k}, x)$ represents the inclusive $D$- or $B$-mesons, 
respectively. The final-state $\psi_f(\Delta {\bf k}, x) = 
a  \psi_M(\Delta {\bf k}, x) + (1-a) \psi_{q\bar{q}}(\Delta {\bf k}, x) $
denotes a superposition of the meson and a dissociated $q\bar{q}$
pair. The survival probability is given by $a^2$ if the light cone
momentum distributions are normalized to unity. We readily obtain:
\begin{eqnarray}
&& \hspace*{-0.7cm} P_s (\chi\mu^2 \xi) 
= \left| \int d^{2}\Delta {\bf k} dx \,
\psi_{f}^* (\Delta {\bf k},x)\psi_{0}(\Delta {\bf k}, x) \right|^{2} 
\nonumber \\
&& \hspace*{-.3cm} = \left| \int d^{2}\Delta {\bf k} dx \,   
{\rm Norm}^2 \left[ \frac{x(1-x)\Lambda^2}
{\chi \mu^{2}\xi+x(1-x)\Lambda^{2} } \right]^{\frac{1}{2}}
\right. \nonumber \\
&& \left. \hspace*{0.1cm} \times 
e^{-\frac{\Delta {\bf k}^{2}}{8}
\left[ \frac{2x(1-x)\Lambda^{2}+\chi \mu^{2} \xi}
{(x(1-x)\Lambda^{2})(\chi \mu^{2}\xi+x(1-x)\Lambda^{2})} \right] }
\,  e^{-\frac{ m_{1}^{2}(1-x)+m_{2}^{2}x } 
{ x(1-x)\Lambda^{2} } } 
 \right|^{2} \nonumber \\
&=& \left|  \int dx \; {\rm Norm}^2 \, 4\pi x(1-x) \Lambda^2
\,  e^{-\frac{ m_{1}^{2}(1-x)+m_{2}^{2}x }{ x(1-x)\Lambda^{2} } } 
\right. \nonumber \\
&&    \left.  \hspace*{-0.2cm}  
\times \left[ \frac{ 2 \sqrt{ x(1-x)\Lambda^{2}}
\sqrt{\chi\mu^{2}\xi+x(1-x)\Lambda^{2}} }
{ \sqrt{x(1-x)\Lambda^{2}}^{2}  + \sqrt{\chi\mu^{2}\xi+x(1-x)
\Lambda^{2}}^2 } \right] \; \right|^2 \, . \;\; \quad 
\label{sprob}
\end{eqnarray}
Eq.~(\ref{sprob}) is one of the main theoretical results derived 
in this Letter. Although the integral over the heavy quark light
cone momentum fraction $x$ has to be taken numerically, direct
comparison of the integrand to Eq.~(\ref{phiofx})  shows that 
the meson survival probability 
$P_s (\chi\mu^2 \xi)  \leq 1$. Equality is reached when 
$\chi \mu^2 \xi = 0$, i.e. in the absence of interactions 
in the medium. We finally note that typical values of 
$\xi$ can be estimated from the requirement 
$b \mu \ll 1$: for $b \mu  = 0.25 - 0.1$ we 
find $\xi \simeq 2 - 3$, respectively.

\section{Heavy meson suppression phenomenology} 

We are now ready to calculate the suppression of heavy hadrons,
$H(c)$ and $H(b)$,  from collisional interactions in the QGP. 
The fragmentation time, averaged over the final-state mesons and baryons 
is calculated as follows:
\begin{eqnarray} 
\frac{1}{\langle \tau_{\rm form}(p_T, t) \rangle } &=& 
\bigg[\sum_i  \int_0^1 dz \, D_{H_i/Q}(z)  \nonumber \\
&& \times \tau_{\rm form}(z,p_T,m_Q, t) \, \bigg]^{-1}  \;.
\label{avtform}
\end{eqnarray} 
In Eq.~(\ref{avtform}) the fragmentation functions are normalized 
to the fragmentation fractions for $Q \rightarrow H(Q)$, which in 
the QCD factorization approach are assumed to be universal,  
$\int_0^1 dz \, D_{H_i/Q}(z) = f_{H_i}(Q)$, 
$\sum_i f_{H_i}(Q) = 1$.  The dissociation time, on the other hand, 
is externally driven by the dynamics of the evolving bulk medium 
and calculated by taking the logarithmic derivative of the 
dissociation probability,
\begin{equation}
P_d(p_T,m_Q,t) = 1 - P_s(p_T,m_Q,t) \; , 
\end{equation}
derived in Eq.~(\ref{sprob}), as follows:
\begin{equation} 
\frac{1}{\langle \tau_{\rm diss}(p_T, t) \rangle} = 
\frac{\partial}{\partial t} \ln  P_d(p_T,m_Q,t) \; .  
\label{avtdiss}
\end{equation} 
We use the same initial soft gluon rapidity density $dN^g/dy$ as
in the calculation of the $\pi^0$ quenching~\cite{Vitev:2005he}
in central Au+Au and Cu+Cu collisions at RHIC and central
Pb+Pb collisions at  LHC and the cumulative momentum transfer
is given by
\begin{equation}
\chi \mu^2 \xi = \beta_Q \frac{\mu^2_0}{\lambda_0} 
\xi \ln \frac{t}{t_0} \;.
\label{cumtr} 
\end{equation}
In Eq.~(\ref{cumtr}) $t_0 \leq t$ and $t_0 = 0.6$~fm,
consistent with QGP formation times used in hydrodynamic simulations 
of bulk observables~\cite{Hirano:2001eu}, and $\beta_Q = dz/dt$.  
Let us denote by
\begin{eqnarray}
\label{convent1}
f^{Q}({p}_{T},t)&=& \frac{d\sigma^Q(t)}{dy d^2p_T} \;,  
 \;   f^{Q}({p}_{T},t=0) = 
\frac{d \sigma^Q_{PQCD}}{dy d^2p_T} \;, \qquad          \\ 
f^{H}({p}_{T},t)&=& \frac{d\sigma^H(t)}{dyd^2p_T} \;,
 \;   f^{H}({p}_{T},t=0) = 0 \; ,   
\label{convent2}
\end{eqnarray}
the double differential cross sections for the heavy quarks and
hadrons (mesons+baryons). Initial conditions are also specified 
above, in particular the heavy quark distribution is given by 
the perturbative $c$-  and $b$-quark jet cross section.
The fragmentation fraction of $b$-quarks into $B_c$-mesons is 
very small. In our work it is neglected and the rate 
equations that describe the competition between $b$- and $c$-quark
fragmentation and $D$- and $B$-meson dissociation decouple
for different heavy quark flavors.  Including the loss and 
gain terms we obtain:
\begin{eqnarray}
\label{rateq1}
\partial_t f^{Q}({p}_{T},t)&=& 
-  \frac{1}{\langle \tau_{\rm form}(p_T, t) \rangle} f^{Q}({p}_{T},t) 
\nonumber \\
&& \hspace*{-2cm}  + \, \frac{1}{\langle 
\tau_{\rm diss}(p_T/\bar{x}, t) \rangle}
\int_0^1 dx \,  \frac{1}{x^2} \phi_{Q/H}(x)  
f^{H}({p}_{T}/x,t) \;, \qquad \\
\partial_t f^{H}({p}_{T},t)&=& 
-  \frac{1}{\langle \tau_{\rm diss}(p_T, t) \rangle} f^{H}({p}_{T},t) 
\nonumber \\
&& \hspace*{-2cm} +\, \frac{1}{\langle 
\tau_{\rm form}(p_T/\bar{z}, t) \rangle}
\int_0^1 dz \,  \frac{1}{z^2} D_{H/Q}(z) 
 f^{Q}({p}_{T}/z,t) \;. \qquad 
\label{rateq2}
\end{eqnarray}
In Eqs.~(\ref{rateq1}) and (\ref{rateq2}) $\bar{z}$ and $\bar{x}$ are
typical fragmentation and dissociation momentum 
fractions and we have checked that
in the absence of a medium, $\tau_{\rm diss}(p_T, t) \rightarrow \infty$,
we recover the PQCD spectrum of heavy hadrons from vacuum 
jet fragmentation.

\begin{figure}[t!]
\begin{center} 
\psfig{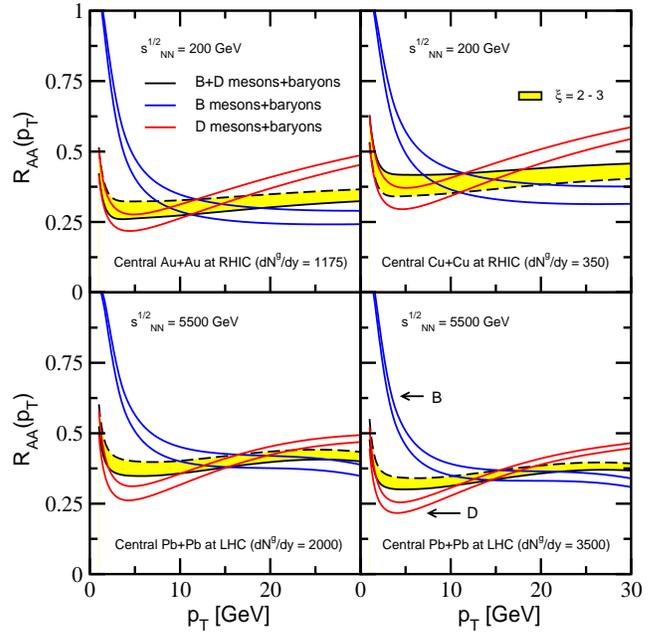}
\caption{ Suppression of $D$- and $B$-meson production via
collisional dissociation in the QGP. Top panels show numerical
results for $R_{AA}(p_T)$ in central Au+Au and Cu+Cu collisions 
at RHIC for gluon rapidity densities $dN^g/dy = 1175$ and  $350$,
respectively~\cite{Vitev:2005he}.
Bottom panel shows predictions for central Pb+Pb collisions at  
LHC for $dN^g/dy = 2100$ and 3500.}
\label{hadquench}
\end{center} 
\end{figure}


We solve the rate equations, Eqs.~(\ref{rateq1}) and 
(\ref{rateq2}), numerically using the initial
 conditions, Eqs.~(\ref{convent1}) and (\ref{convent2}), and
obtain the final spectra of heavy hadrons at 
$t \gg L_T^{QGP}, \tau_{\rm form}$  and $\tau_{\rm diss}$. 
The nuclear modification factor, $R_{AA}(p_{T})$, which arises 
from the collisional dissociation of  $D$-and $B$-mesons, 
is shown in Fig.~\ref{hadquench} for  physical situations 
expected to be prevalent at RHIC and LHC. In spite of the
inherently different physics mechanisms that drive the 
suppression of light and heavy hadrons, comparison to the 
quenching of the $\pi^0$~\cite{Vitev:2005he} can be done 
if the same model for the QGP properties, such as the gluon 
rapidity density $dN^g/dy$, is used.  
The theoretical uncertainty comes from varying the parameter 
$\xi$ in its natural range and $\mu_{0}$ and $\lambda_{0}$ 
are set by the overall gluon multiplicity and the QGP 
formation time, $\tau_0$, using thermodynamic arguments.
The top panels show results for central Au+Au and Cu+Cu collisions 
at the maximum RHIC energy $\sqrt{s_{NN}} = 200$~GeV. 
The bottom  panels  present $R_{AA}(p_T)$ for two different $dN^g/dy$
extrapolations in  central Pb+Pb collisions at  LHC.

One important feature of this approach is the sensitivity to 
the build-up of the QGP density. If $\tau_0 << \tau_{\rm form}$ 
and $\rho(t)$ decreases rapidly, hadrons would not have formed
at times when the dissociation mechanism is most efficient. This 
leads to reduced suppression, which can be seen in the high-$p_T$ 
behavior of the $D$-meson $R_{AA}$ and understood when one recalls 
that $\tau_{\rm form} \propto p_T$.  
Since the distortion of the  
light cone wave functions reflects the accumulated squared 
transverse momentum transfer, the reduced suppression at 
$p_T \leq m_Q $ arises from the velocity factor $\beta_Q$ in
Eq.~(\ref{cumtr}). We observe in Fig.~\ref{hadquench} that 
with initial formation time $\tau_0 = 0.6$~fm, consistent      
with the one used in hydrodynamic simulations to
describe bulk QGP observables~\cite{Hirano:2001eu}, the 
moderate-  and high-$p_T$ suppression of heavy hadron production 
is comparable to that of lighter hadrons. We have checked that
for QGP formation time $\tau_0 = 0.2$~fm the $p_T >  5$~GeV 
heavy quark quenching
is reduced by a factor of $\sim 1.5$.   Finally, one should  
note that  dissociation/fragmentation both emulate energy loss 
by shifting the quarks/hadrons to lower transverse momenta. For 
example, $B$- and $D$-meson attenuation is sensitive to the 
partonic slope. For this reason the suppression at  LHC is 
found to be comparable or slightly smaller than the one calculated
at RHIC in spite of the larger QGP densities and temperatures.

Contrary to calculations that emphasize radiative and collisional
heavy quark energy 
loss~\cite{Djordjevic:2005db,Armesto:2005mz,Wicks:2005gt,vanHees:2004gq}, 
QGP-induced dissociation predicts $B$-meson suppression comparable to
or larger than that of $D$-mesons at transverse momenta as low as 
$p_T \sim 10$~GeV, see Fig.~\ref{hadquench}.  This is  due to the 
significantly smaller formation times for $H(b)$ relative to $H(c)$.
Thus, each fragmentation/dissociation cycle proceeds at a much 
faster rate for $b$-quarks/$B$-mesons. This is an example where
the large mass facilitates the hadron suppression mechanism. 
The observable effect of this faster rate in the 
quenching of  the final hadron distributions is amplified  by the 
significance of the early $t \sim \tau_0 $ hot and dense stage 
in the dynamical 
evolution  of the QGP. Our simulations show that the rates 
in Eqs.~(\ref{rateq1}) 
and (\ref{rateq2}) may  play a more important role than the effective
fractional energy losses $\epsilon = \Delta E /E \approx 1- \bar{z}$ 
and $\epsilon  \approx 1-\bar{x}$, which are larger for 
$c$-quarks/$D$-mesons.

\begin{figure}[t!]
\begin{center} 
\psfig{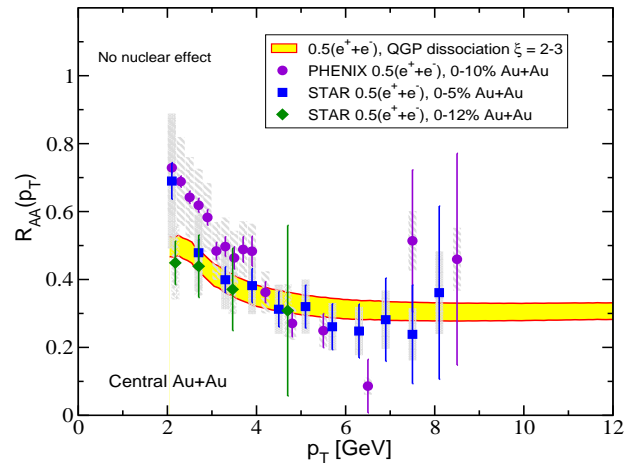}
\caption{ Suppression of inclusive non-photonic electrons  from
$D$- and $B$-meson spectra softened by collisional dissociation 
in central  Au+Au collisions.  Data on non-photonic electron 
quenching from  PHENIX~\cite{Adler:2005xv,Ralf} and 
STAR~\cite{Abelev:2006db,Alex} is also shown.}
\label{elquench}
\end{center} 
\end{figure}

Inclusive non-photonic electron data for the most central 
Au+Au collisions at $\sqrt{s_{NN}}=200$ GeV from 
PHENIX~\cite{Adler:2005xv} and STAR~\cite{Abelev:2006db} 
are compared to the current calculation in Fig. \ref{elquench}.  
We used the PYTHIA event generator~\cite{Sjostrand:2006za} 
to  simulate the $(e^+ + e^-)/2$ spectra  coming from the 
primary decays of the $D$- and $B$-mesons in our baseline p+p 
and QGP-modified A+A results. We find that the single electron
suppression can be as large as a factor of four and approximates
well the quenching extracted from available 
data~\cite{Adler:2005xv,Ralf,Abelev:2006db,Alex}.  
We  emphasize that such agreement between theory and experiment 
is not achieved at the cost of neglecting the contribution of 
the $B$-mesons to the non-photonic electron spectra.


\vspace*{0.5cm}

\section{Conclusions} 

It has been recently suggested that collisional dissociation 
of heavy quarkonia in the quark-gluon plasma~\cite{Wong:2004zr} 
may be a possible explanation for the suppression of their 
production rate in nucleus-nucleus collisions. It is thus 
surprising that until now a similar physics mechanism has
not been considered for open heavy flavor. 
In this Letter, we investigated the perturbative QCD dynamics 
of open charm and beauty production and, in the framework of 
the reaction operator (GLV) 
approach~\cite{Gyulassy:2000fs,Gyulassy:2000er,Gyulassy:2002yv,Qiu:2003pm}
extended to composite $q\bar{q}$ systems, derived the 
medium-induced dissociation probability for heavy $D$- and 
$B$-mesons  traversing dense nuclear matter. We showed that 
the effective energy loss, which arises from the sequential 
fragmentation and dissociation of heavy quarks and mesons, is 
sensitive to the interplay between the formation times of the hadrons 
and the QGP and the detailed expansion dynamics of hot nuclear matter. 
The proposed new attenuation mechanism, which stems from 
the short formation times of $D$- and $B$-mesons and underlies 
the suppression  of the inclusive non-photonic 
electrons~\cite{Adler:2005xv,Ralf,Abelev:2006db,Alex} in nucleus-nucleus 
collisions at RHIC, was found to be compatible with the measured 
large, factor of four to five, quenching  for heavy flavor.

Previous studies, based on radiative and collisional parton energy 
loss~\cite{Djordjevic:2005db,Armesto:2005mz,Wicks:2005gt}
and heavy quark diffusion~\cite{vanHees:2004gq} under-predict the 
suppression of non-photonic electrons in central Au+Au collisions
due to the small $b$-quark quenching. A natural consequence of 
the approach, presented in this Letter as a viable alternative to 
existing calculations, is that $B$-mesons are attenuated as much as
$D$-mesons at transverse momenta as low as $p_T \sim 10$~GeV. 
While we anticipate that a comparable description of the attenuation
of the non-photonic electrons may be achieved when partonic energy 
loss is combined with  quark-resonance interactions near the 
QCD phase transition in Langevin transport simulations, the 
hierarchy $R_{AA}(H(c)) \ll  R_{AA}(H(b))$ will not be changed 
in the accessible transverse momentum range~\cite{prep}.   
We conclude that  robust experimental determination  of the dominant 
mechanism for in-medium modification of open heavy flavor would 
require direct and separate measurements of the $B$- and $D$-meson  
$R_{AA}$ distributions versus $p_T$ and centrality in 
collisions of heavy  nuclei.

\vspace*{-0.1in}   

\begin{acknowledgments}
Illuminating discussion with Y. Akiba, T. Goldman, P. B. Gossiaux, 
H. van Hees, C.-M. Ko, D. Molnar, R. Rapp and N. Xu is 
gratefully acknowledged. 
This work is supported in part by the United  States Department of Energy 
under Contract No. DE-AC52-06NA25396, Grant No. DE-FG02-93ER40764 and
the J. Robert Oppenheimer Fellowship of the Los Alamos National Laboratory.
\end{acknowledgments}

\end{document}